%% file: main.tex
\documentclass{article}

\usepackage{PRIMEarxiv}
\usepackage{algpseudocode}
\usepackage{algorithm, algorithmicx}
\usepackage{amsmath}
\usepackage{caption}
\usepackage[utf8]{inputenc} 
\usepackage[T1]{fontenc}    
\usepackage{float}
\usepackage{biblatex} 
\usepackage{hyperref}       
\usepackage{url}            
\usepackage{booktabs}       
\usepackage{amsfonts}       
\usepackage{nicefrac}       
\usepackage{microtype}      
\usepackage{lipsum}
\usepackage{fancyhdr}       
\usepackage{graphicx}       
\usepackage{listings}
\usepackage{framed}
\usepackage{multirow}
\usepackage{tabularray}
\usepackage{subcaption}

\graphicspath{{media/}}     
  
\title{HadaCore: Tensor Core Accelerated Hadamard Transform Kernel
}

\author{
  Krish Agarwal$^*$ \\
  University of Texas at Austin \\
  \texttt{krishagarwal@utexas.edu} \\
   \And
  Rishi Astra$^*$ \\
  University of Texas at Austin \\
  \texttt{raa3897@utexas.edu} \\
   \And
  Adnan Hoque \\
  IBM T.J. Watson Research Center \\
  Yorktown Heights, NY, United States\\
  \texttt{adnan.hoque1@ibm.com} \\
   \And
  Mudhakar Srivatsa \\
  IBM T.J. Watson Research Center \\
  Yorktown Heights, NY, United States\\
  \texttt{msrivats@us.ibm.com} \\
   \And
  Raghu Ganti \\
  IBM T.J. Watson Research Center \\
  Yorktown Heights, NY, United States\\
  \texttt{rganti@us.ibm.com} \\
   \And
  Less Wright \\
  Meta AI \\
  \texttt{less@meta.com} \\
   \And  
   Sijia Chen \\
  Meta AI \\
  \texttt{sijiac@meta.com} \\
}

\begin{document}
\maketitle

\def\thefootnote{*}\footnotetext{Equal contribution.}
\def\thefootnote{**}\footnotetext{Work done at IBM Research using IBM facilities. IBM has released the source code.}
\def\thefootnote{\arabic{footnote}}

\begin{abstract}

We present HadaCore, a modified Fast Walsh-Hadamard Transform (FWHT) algorithm optimized for the Tensor Cores present in modern GPU hardware. HadaCore follows the recursive structure of the original FWHT algorithm, achieving the same asymptotic runtime complexity but leveraging a hardware-aware work decomposition that benefits from Tensor Core acceleration. This reduces bottlenecks from compute and data exchange. On Nvidia A100 and H100 GPUs, HadaCore achieves speedups of 1.1--1.4x and 1.0--1.3x, with a peak gain of 3.5x and 3.6x respectively, when compared to the existing state-of-the-art implementation of the original algorithm. We also show that when using FP16 or BF16, our implementation is numerically accurate, enabling comparable accuracy on MMLU benchmarks when used in an end-to-end Llama3 inference run with quantized (FP8) attention.\footnote{Our code is publicly available at \url{https://github.com/pytorch-labs/applied-ai/tree/main/kernels/cuda/inference/hadamard_transform}}
\end{abstract}


\section{Introduction}

QuaRot \cite{ashkboos2024quarotoutlierfree4bitinference} and SpinQuant \cite{liu2024spinquantllmquantizationlearned} both propose methods to increase the numerical accuracy of INT4 and INT8 quantization in LLMs. These methods rotate model weights and activations since a rotation is statistically likely to reduce the magnitude of outliers, as it “distributes” extreme values among other (less extreme) dimensions, and rotation is also an easily invertible operation using the inverse of a rotation matrix. These methods can also improve FP8 inference accuracy, such as in FlashAttention-3 \cite{shah2024flashattention3fastaccurateattention}.

\begin{figure}[ht]
    \centering
    \includegraphics[width=\linewidth]{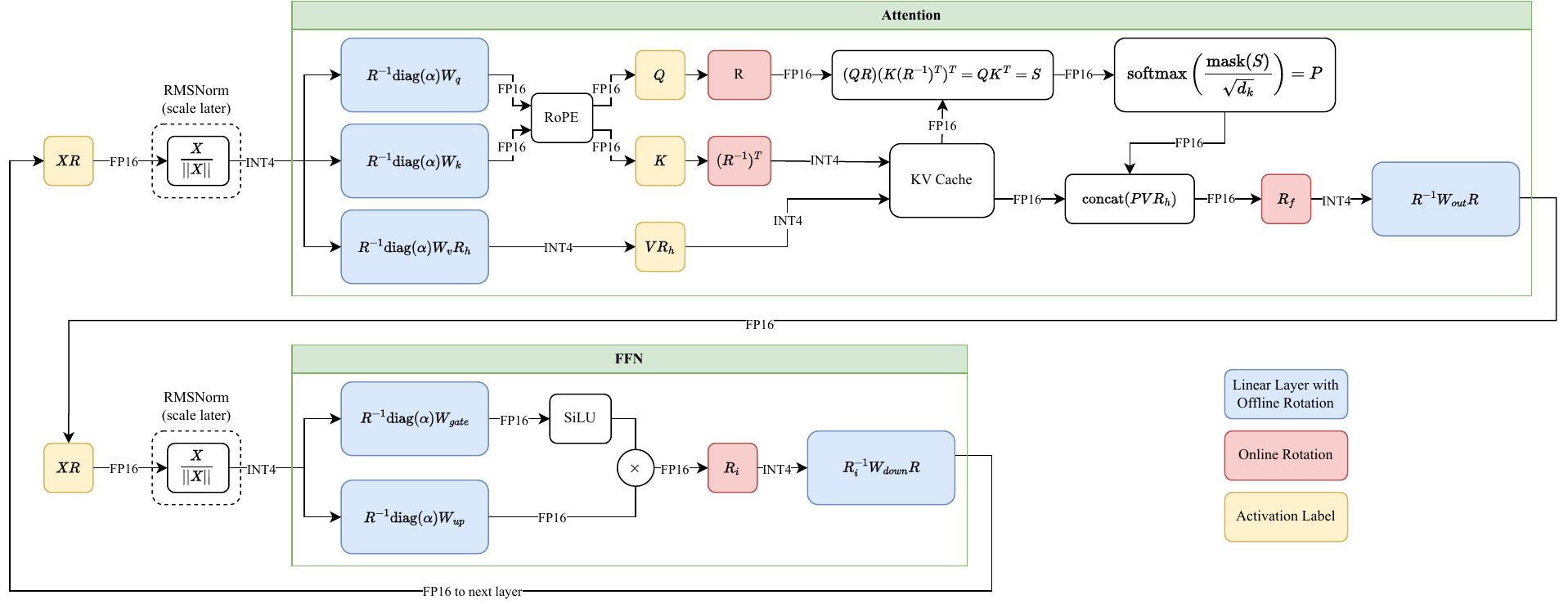}
    \caption{Transformer block showing online (red) and offline rotations (blue) in QuaRot.}
    \label{fig:quarot_diagram}
\end{figure}

Placing these rotation matrices into the model runtime introduces extra overhead. Naively, these rotations would be full matrix multiplications. For Llama-2 7B, if we used INT8 quantization, we might expect a theoretical 2x speedup compared to FP16 or BF16 inference, but using FP16  or BF16 matmuls for these rotations could increase the linear layer computation to 110\% of the original FP16 or BF16 model (supposing INT8 is half the cost).

QuIP\# \cite{tseng2024quipbetterllmquantization} proposes the use of Walsh-Hadamard matrices, a special type of rotation matrix that can be applied much faster than a matrix multiplication (matmul) using the Fast Walsh-Hadamard Transform algorithm. QuaRot and SpinQuant also opt to use these special rotations. Our work focuses on improving the performance of this transform. 

\section{Background}

\subsection{Hadamard Matrices}

Hadamard matrices are square, orthogonal matrices with values exclusively either 1 or -1 ($\pm\frac{1}{\sqrt{d}}$ for a $d$-size Hadamard matrix when normalized). Walsh-Hadamard matrices are power-of-2 size Hadamard matrices which can be constructed recursively. Because of this recursive construction, multiplying an $m\times n$ activation matrix $x$ with an $n$-sized Walsh-Hadamard matrix $H$ can be performed in $O(mn\log{(n)})$ time as opposed to the $O(mn^2)$ time if general matrix multiplication was used. Furthermore, this algorithm, called the Fast Walsh-Hadamard Transform (FWHT), does not store $H$ in memory, reducing matmul memory bottlenecks.

\subsection{Fast Walsh-Hadamard Transform Algorithm}

The Fast Walsh-Hadamard Transform algorithm \cite{fastwalshhadamardtransform} uses a recursive formula to apply an $n$-size Walsh-Hadamard transform to a vector/matrix $x$ given the $x$ already transformed by an $(n/2)$-sized Hadamard transform. This is very related to Sylvester’s construction of Hadamard matrices \cite{Sylvester1867LXTO}, except that the Hadamard transform is applied without materializing the corresponding matrix.

\input{code/fwht}

In the above code, the outer loop represents the recursive iteration (the current Hadamard transform size is $2h$). The inner loop completes the current Hadamard transform size, with the middle loop repeating this to span all elements (e.g. if our vector is of size 8 or matrix of size $n \times 8$, the size-2 Hadamard would be repeated 4 times to span all elements, of size 4 repeated twice, of size 8 repeated once).

This algorithm has roughly $m\log_2{(n)} \cdot n$ floating point operations when applied to an $m\times n$ matrix, making it $O(mn\log(n))$.

\subsection{Parallelization and Running on the GPU}

In the base case of applying a size-2 Hadamard transform, work needs to be done on every pair of adjacent elements. Then in the next iteration, to accomplish a $4\times 4$ Hadamard transform, work needs to be done on every pair of elements separated by 2 (i.e. 0 and 2, 1 and 3, 4 and 6, 5 and 7, etc.). This repeats for higher Hadamard sizes by nature of the algorithm. Importantly, each iteration depends on the output from the previous iteration.

The elements accessed in the inner code do not overlap between iterations of the inner and middle loop. This makes it simple to parallelize the 2 inner loops, which together span all $m$ elements of the $x$ vector/matrix, on a GPU. If $x$ is a matrix, we can also parallelize across $n$ rows. However, we cannot parallelize the outermost loop due to each iteration depending on the previous.

Additionally, there are synchronization challenges. Suppose that, per iteration, each thread does work on a single pair of elements. That thread will not be working on the same pair in the next iteration, which means it must have access to the results from another thread’s work in this iteration to use in the next iteration.

Modern GPUs group threads into warps of 32 threads and further into threadblocks (CTAs) of up to 1024 threads. If each thread processes 2 elements (with the simple parallelization previously explained), we can process up to 64 elements (iteration 5) by exchanging data in the warp and up to 2048 (iteration 11) elements by exchanging within a threadblock through shared memory. If our Hadamard transform size is greater, each thread must process more than 2 elements, or expensive cross-threadblock synchronization and global memory are required.

\subsection{\texttt{fast-hadamard-transform} Library}

The Dao AI lab provides an optimized CUDA \texttt{fast-hadamard-transform} library \cite{tridaofht} that implements the Fast Walsh-Hadamard Transform algorithm.

The library performs a right-Hadamard transform, parallelizing across rows of the input matrix through the launch grid and within each row by using up to 256 threads per row. The library processes 8 elements per thread (using fewer elements per thread was probably not worth the extra data movement and synchronization). We can interpret a row processed by a single threadblock to be shaped as \texttt{(n\_chunk, warps\_per\_threadblock, threads\_per\_warp, 8)}.
\begin{itemize}
    \item Each thread handles \texttt{n\_chunk} chunks of 8 elements each (each 8 elements are contiguous, all threads’ first chunks are contiguous together, then all second chunks, etc.)
    \item First process each chunk “owned” by individual threads (no syncing)
    \item Exchange elements within each warp and process them to carry out the next few iterations (no explicit synchronization required)
    \item Synchronize across the threadblock and transpose data so each warp has the data it needs for the next iterations
    \item Process elements within each warp again
    \item Synchronize and transpose data back
    \item Have each thread transpose its data so it has 8 chunks of \texttt{n\_chunk} elements (as opposed to the other way around) and process each chunk within each thread (no syncing)
\end{itemize}
These optimizations provide 15 iterations of the algorithm with only 2 threadblock syncs, allowing up to a $2^{15}$-size Hadamard transform.

\section{Method}
Modern Nvidia GPUs have specialized hardware called Tensor Cores \cite{tensorcore}, which can compute matrix multiplications on matrix sizes such as $16\times 16$ about 8x faster than CUDA cores (general purpose GPU hardware). Using these Tensor Cores would therefore boost our performance. However, the original fast Hadamard transform algorithm is explicitly an alternative to using matrix multiplication, and thus it cannot be directly run on Tensor Cores.
The \texttt{mma} instruction can perform $16\times 16$ by $16\times 8$ matrix multiplication, so we can use 2 such operations to perform a $16\times 16$ by $16\times 16$ matrix multiplication. We can therefore perform a 16-size Hadamard transform on a $16\times 16$ matrix chunk using these 2 Tensor Core \texttt{mma} operations. Furthermore, we can use inline assembly (PTX) to use the Tensor Cores with data already in registers.

\subsection{Hadamard Sizes up to 256}

Naively, if we were to integrate the above into the original Fast Walsh-Hadamard Transform algorithm, we could treat a size-16 Hadamard transform as a base case as opposed to size-2 Hadamard being the base case, for which we would use a Tensor Core. Just by using this to replace the first 4 outer iterations of the regular algorithm, this new algorithm takes $Cmn\log_2{\left(\frac{n}{16}\right)}$ time, where $C$ is the latency of the 2 Tensor Core \texttt{mma} instructions. 

Although a $\frac{mn}{16}\times 16$ by $16\times 16$ matmul using Tensor Cores instead of four $\frac{mn}{2}\times 2$ by $2\times 2$ matmuls uses 2x more floating-point operations for the same result, Tensor Cores have \textasciitilde 8x the FLOPS and reduce the number of iterations thus reducing threadblock/global synchronizations.

To extend this, we can rearrange the elements such that applying the base case again achieves the next 4 iterations. First, we can apply the 16-size Hadamard to a $1\times 256$ row vector chunk by reshaping it to $16\times 16$. This effectively applies the 16-size Hadamard to groups of 16 elements in the row. Then to rearrange elements, we can transpose the $16\times 16$ result and apply the $16\times 16$ base case again (now acting on elements 16 apart due to the transpose). Finally, if we transpose the result back and view it again as a $1\times 256$ row vector, we achieve a size-256 Hadamard transform.

\begin{figure}[ht]
    \centering
    \includegraphics[page=1,width=\linewidth]{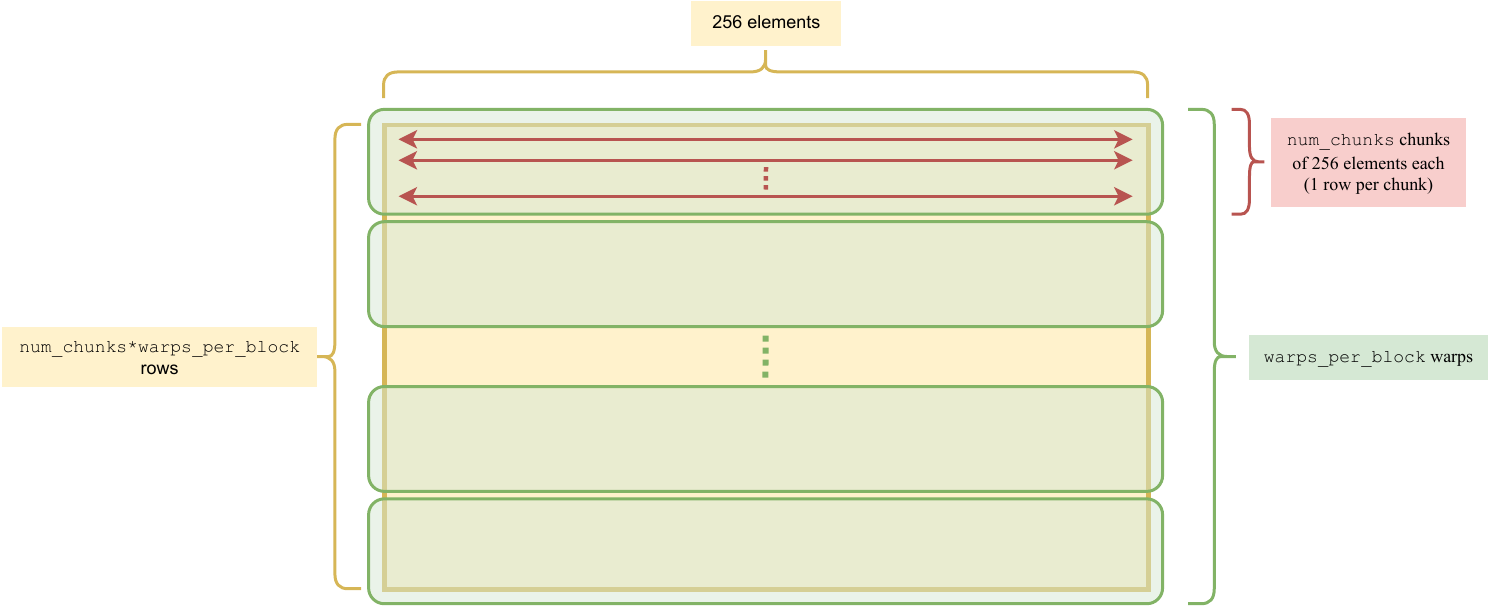}
    \caption{$1\times 256$ vectors (rows) visualized for rotating by a size-256 Hadamard. The batches (number of rows) can be split among warps of the GPU.}
    \label{fig:first_256}
\end{figure}

\subsection{Supporting Sizes Larger than 256}

So far, we would only achieve up to a 256-size Hadamard with the above strategy because every contiguous chunk of $h$ elements must be synced to achieve an $h$-sized Hadamard. Suppose that for a $1\times n$ row vector we are trying to do an $n$-size Hadamard transform, where $n = 2^k > 256$. We can first view this as a $\frac{n}{256}\times 256$ matrix and apply a 256-size Hadamard transform to $\frac{n}{256}$ separate chunks with the method described above. To go beyond 256, we can transpose the current result to a $256\times \frac{n}{256}$ matrix and apply an $\frac{n}{256}$-sized Hadamard transform, again using Tensor Core \texttt{mma}s, and transpose the result back. However, we would lose parallelism if we were to have a single warp process the full $1\times n$ vector, especially for larger sizes like $2^{15}$.

The approach we use is to process a full $1\times n$ vector per threadblock, where each threadblock has \texttt{warps\_per\_block} warps and each warp processes \texttt{num\_chunks} chunks of 256 elements. We choose \texttt{warps\_per\_block} and \texttt{num\_chunks} such that $256 \cdot \texttt{warps\_per\_block} \cdot \texttt{num\_chunks} = n$. Our strategy is the following:
\begin{enumerate}
    \item Have each warp process its chunks of 256 elements as described above (each 256-element chunk is a row in the $\frac{n}{256}\times 256$ view).
    \item Store the result back in shared memory.
    \item Sync across the threadblock.
    \item Read from shared memory transposed so that each warp has $\frac{256 \cdot \texttt{num\_chunks}}{n / 256}$ columns of $\frac{n}{256}$ elements from the $\frac{n}{256}\times 256$ view (this still amounts to each warp processing \texttt{num\_chunks} chunks of 256 elements each).
    \item Process the new data in chunks of $256$ (where each chunk comprises of one or more columns) by using the same method as up to size $256$ (just like the Dao AI Lab kernel, HadaCore supports up to size $2^{15} = 256 \cdot 128$, so this works out).
\end{enumerate}

\begin{figure}[ht]
    \centering
    \includegraphics[page=2,width=\linewidth]{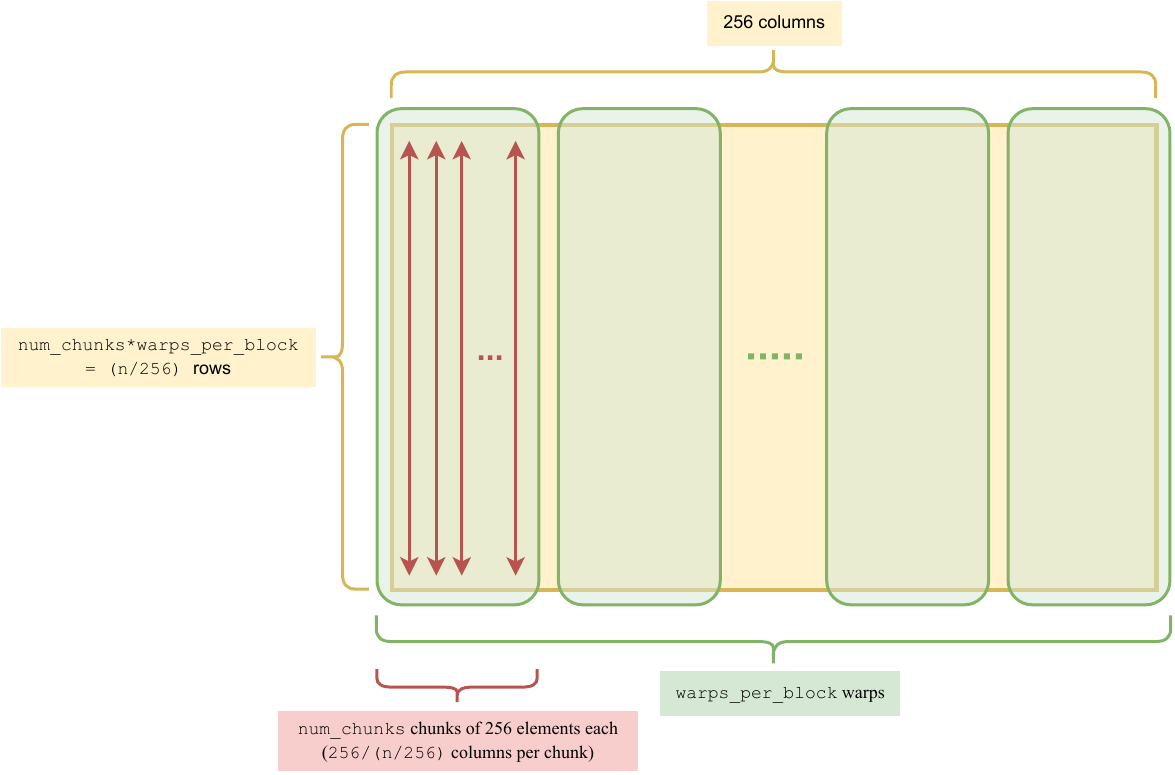}
    \caption{The $1\times 256$ vectors now transposed, allowing us to operate on the factor $\frac{n}{256}$.}
    \label{fig:second_256}
\end{figure}

Loading transposed data means that the loads are uncoalesced. We solve this in the following ways:
\begin{itemize}
    \item For sizes 512, 1024, and 2048, due to the Tensor Core register layout, we are able to simply load each chunk coalesced and then use warp-level shuffle operations to get each thread its own data.
    \item For sizes 4096 and above, we can coalesce our loads by loading all chunks upfront into registers and then doing warp-level shuffle operations to get each thread its own data.
\end{itemize}

\subsection{Non-Power of 16 Sizes}

The proposed algorithm relies on the desired Hadamard size being a power of 16 rather than a general power of 2, but this issue is easily resolved. If the desired Hadamard size is $d \ne 16^n$, we can factorize $d = 2^m \cdot 16^n$, where $0 < m < 4$. The $16^n$-size Hadamard can be achieved with $n$ iterations of applying a 16-size Hadamard with intermediate rearranging as specified above, and the $2^m$ size Hadamard can be achieved in one extra iteration by using an appropriate $16\times 16$ matrix (e.g. for $m = 3$, have an $8\times 8$ Hadamard repeated on the diagonal). In this way, the algorithm is the same as if the Hadamard size was the next power of 16, except that the last $16\times 16$ matrix is a diagonal tiling of a smaller Hadamard.

\subsection{Algorithm Overview}

The original Fast-Walsh Hadamard Transform algorithm achieves a $2^n$-size Hadamard transform given an input already transformed by a $2^{n-1}$-size Hadamard. This is analogous to producing a $2^n$ size Hadamard using the Kronecker product of a $2^{n-1}$-size Hadamard with a size-2 Hadamard. Similarly, by using a 16-size Hadamard per iteration, our algorithm is analogous to using a Kronecker product of a $2^{n-4}$ size Hadamard with a size-16 Hadamard to achieve a $2^n$ size Hadamard transform, without materializing the Hadamard matrix in memory.

Our algorithm requires $\left(\frac{mn}{16}\right)(16\cdot 16)\lceil{\log_{16}{(n)}}\rceil = 16mn \lceil \log_{16}{(n)}\rceil \ge 16mn \log_{16}{(n)} = 4mn \log_2{(n)}$ floating-point operations. Comparatively, the regular algorithm requires $\left(\frac{mn}{2}\right)(2\cdot 2)\log_2{(n)} = 2mn\log_2{(n)}$ floating-point-operations. With at least 2x the floating-point operations, we might expect our algorithm to be slower, but we still expect it to be faster because:

\begin{itemize}
    \item By using Tensor Cores, we have 8x the FLOPS available to us.
    \item By using Tensor Core \texttt{mma} operations, we require less explicit data shuffling, meaning less warp stalls due to synchronization.
    \item This also means we perform far fewer non-Hadamard Transform operations: the \texttt{fast-hadamard-transform} library uses complicated indexing to achieve its warp-level data shuffling, meaning a much higher ALU load; in comparison, our warp-level shuffling is a simple hardware transpose.
    \item Our algorithm is more flexible to varying threadblock sizes, which means we are able to optimize our configurations to achieve higher occupancy compared to the \texttt{fast-hadamard-transform} library (this is especially apparent for a 128-size Hadamard transform as shown in our results section below).
\end{itemize}

\section{Results}

\subsection{Performance Tests}

We compare HadaCore's performance against the Dao AI lab’s \texttt{fast-hadamard-transform} library on both an A100 and an H100. We measure our relative speeds across different Hadamard sizes, and we also vary the element count (analogous to number of input matrix rows). We chose to distinguish element count instead of row count since runtimes for different Hadamard sizes are most similar for the same amount of data rather than the batch size. The graphs and tables of our results are available in Appendix \ref{appendix:graphs}.

Here are some notes about our results:
\begin{itemize}
    \item We have a noticeable increase in performance for an element count of 8M on the A100 (16M on the H100) from an optimization we performed with using in-place rotations. This is explained in further detail in Appendix \ref{appendix:inplace}.
    \item Using a Hadamard size of 512 and a small element count has noticeably worse speedup compared to our other results. 512 is the smallest size that requires more than one 256-size fragment to be synchronized. This means a Hadamard size of 512 incurs roughly the same overhead that all Hadamard sizes above 256 incur from threadblock syncing and shared memory data shuffling. While the \texttt{fast-hadamard-transform} CUDA kernel also performs the same threadblock sync above a size of 256, our shared memory data shuffling is likely more expensive due to adhering to Tensor Core register layouts. Additionally, our implementation incurs the full cost of a $16\times 16$ matmul (a $2\times 2$ Hadamard is tiled along the diagonal) while the Dao AI Lab’s implementation only incurs an extra outer iteration of the general algorithm.
    \item Similarly, Hadamard size 8K has a relatively lower speedup since it requires the full 4 iterations of $16\times 16$ matmuls ($16^3$ = 4K < 8K), the same amount as 32K, while 4K would require 3. Again, the \texttt{fast-hadamard-transform} library only incurs the cost of an extra outer iteration to go from 4K to 8K.
    \item With our coalescing scheme for loading transposed data from shared memory, we don’t necessarily achieve full coalescing for Hadamard sizes 8K and above. For sizes 8K, 16K, and 32K, we require each warp to process 8, 16, and 32 chunks respectively for full coalescing. However, this may sacrifice parallelism, so we empirically select configurations for each of these Hadamard sizes. This results in varying runtimes between sizes that use the same amount of computation, such as 8K/16K/32K.
    \item The H100 results are overall worse than the A100 results. This is likely due to architectural differences, a different compute/bandwidth ratio, and different loading instructions. We focused on pre-Hopper GPUs during development.
\end{itemize}

\subsection{End-To-End Tests}

Beyond basic unit tests that check the output of HadaCore against the output of an explicit Hadamard matrix multiplication, we analyze MMLU \cite{hendrycks2021measuringmassivemultitasklanguage} accuracy for Llama-3.1 8B to compare implementations that use FP8 attention with and without rotations. These tests validate the usefulness of Hadamard rotations for reducing quantization error as well as demonstrate that HadaCore does not sacrifice numerical accuracy by being faster.

\begin{table}[H]
\centering
\begin{tblr}{
  width = \linewidth,
  colspec = {Q[417]Q[317]Q[198]},
  column{2} = {c},
  column{3} = {c},
  cell{1}{2} = {c=2}{0.515\linewidth},
  cell{2}{2} = {c=2}{0.515\linewidth},
  cell{3}{2} = {c=2}{0.515\linewidth},
  hlines,
  vlines,
}
 & \textbf{Average 5-Shot MMLU Accuracy ($\uparrow$) } & \\
\textbf{FP16 Baseline} & 65.38 & \\
\textbf{FP8 attention (no rotation)} & 64.40 & \\
 & \textbf{Dao AI Lab Kernel} & \textbf{HadaCore}\\
\textbf{FP8 attention (with rotation)} & 65.45 & 65.09
\end{tblr}
\end{table}

The results above show that HadaCore maintains a comparable quantization error reduction as the Dao AI Lab kernel for Hadamard rotations in FP8 attention.

\section{Conclusion}

We showcased our speedups achieved by moving the Fast Walsh-Hadamard transform algorithm into a CUDA kernel that leverages Tensor Core acceleration. HadaCore achieves speedups of around 1.1--1.4x and up to 3.5x the speed of the Dao AI CUDA kernel. Further, by running MMLU benchmarks on Llama-3.1 8B with FP8 attention, we show that rotating with HadaCore achieves similar quantization error reduction as the Dao AI CUDA kernel while providing computational acceleration. In the future, we plan to implement a version of HadaCore for OpenAI Triton \cite{triton}, do more optimizations for the newer H100 Hopper architecture, and experiment with more advanced techniques such as kernel fusion to support fused Hadamard transform and quantization.

\printbibliography

\newpage
\appendix

\section{Graphs and Tables}
\label{appendix:graphs}

\subsection{Graphs}

\begin{figure}[H]
    \centering
    \includegraphics[width=.99\linewidth]{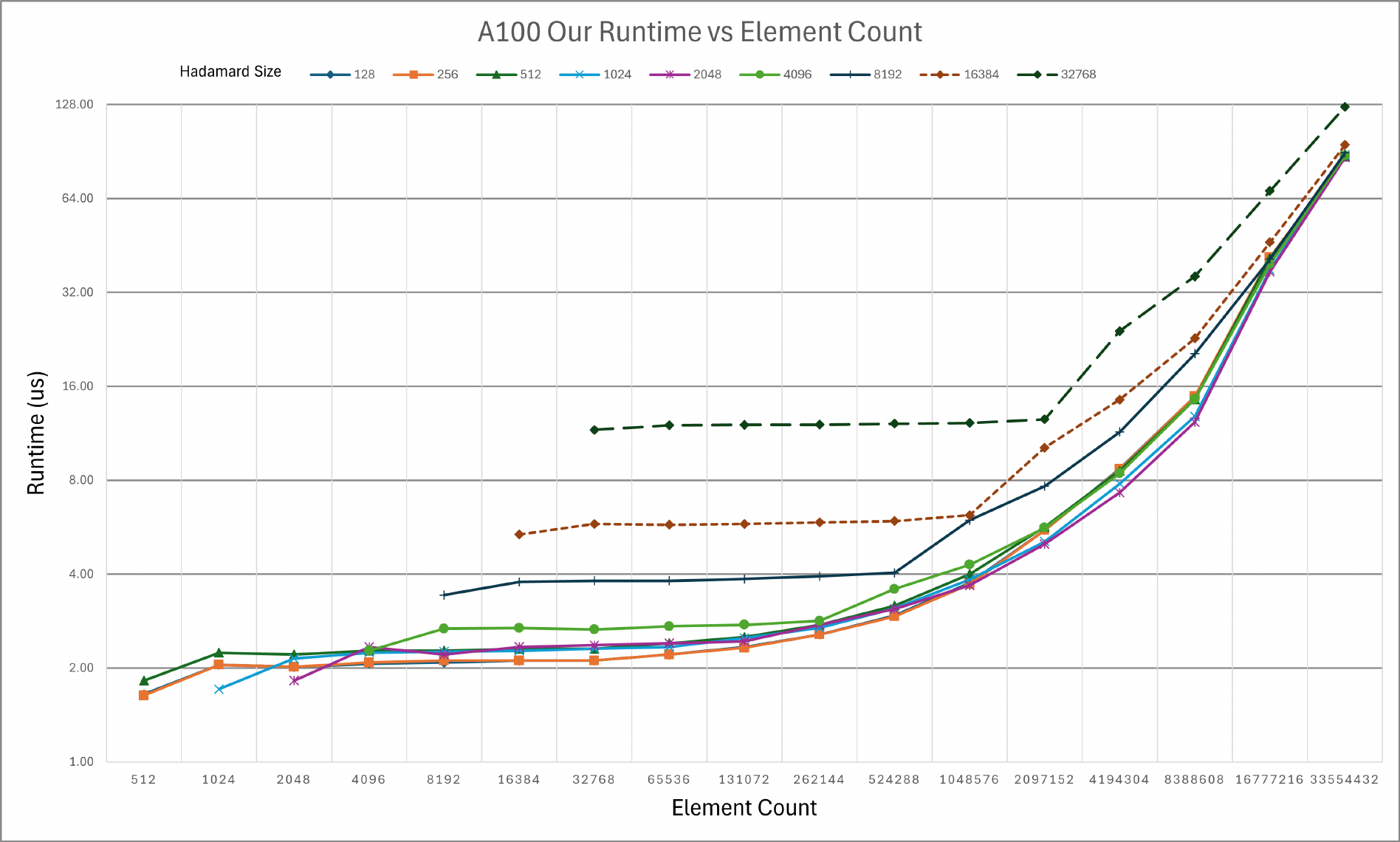}
    \includegraphics[width=.99\linewidth]{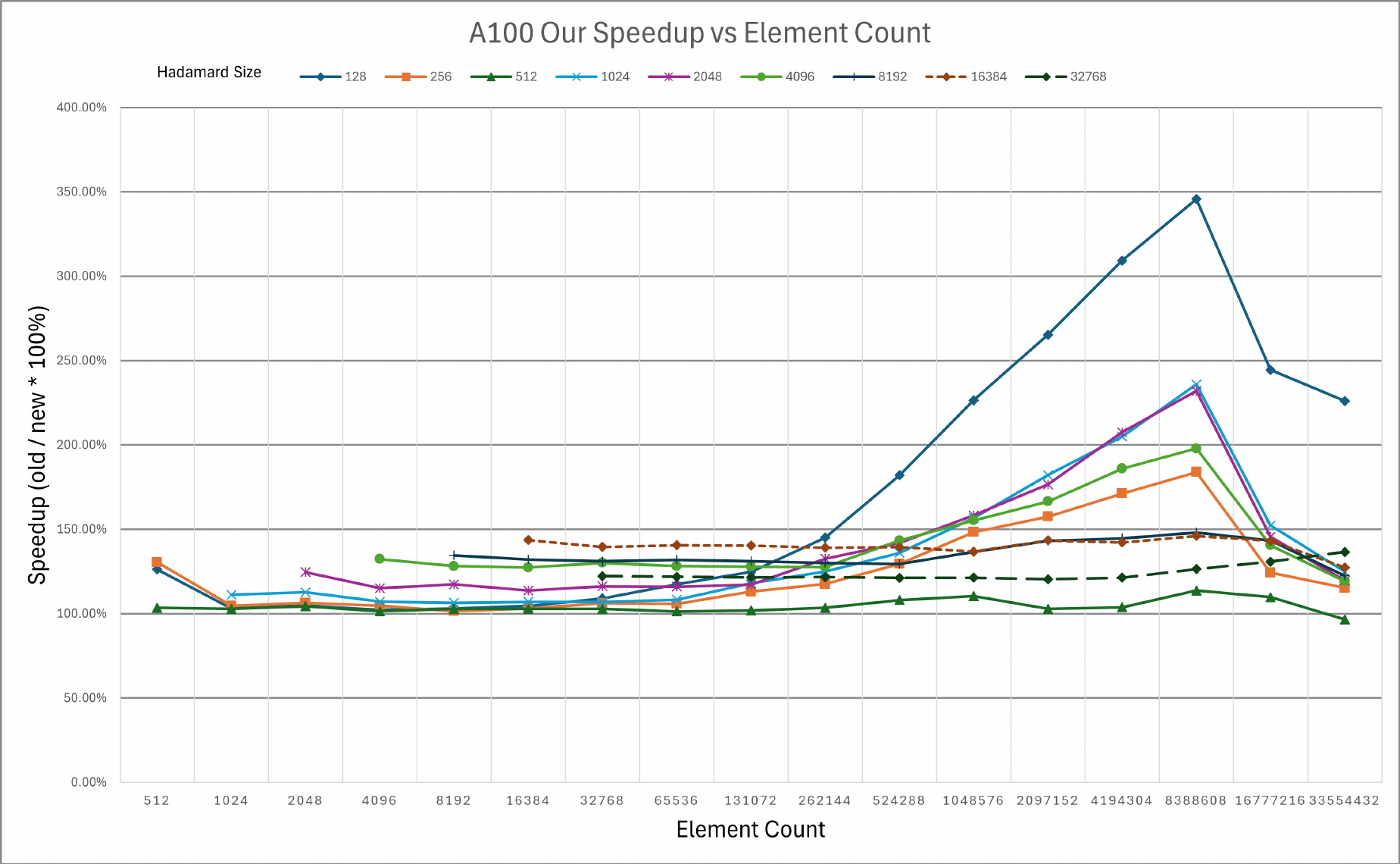}
    \caption{Graphs of runtime and speedup of HadaCore against the Dao AI Lab kernel, measured on an A100-PCIe. Each series represents a different rotation size, and the x-axis is the element count of the input tensor.}
\end{figure}

\begin{figure}[H]
    \centering
    \includegraphics[width=\linewidth]{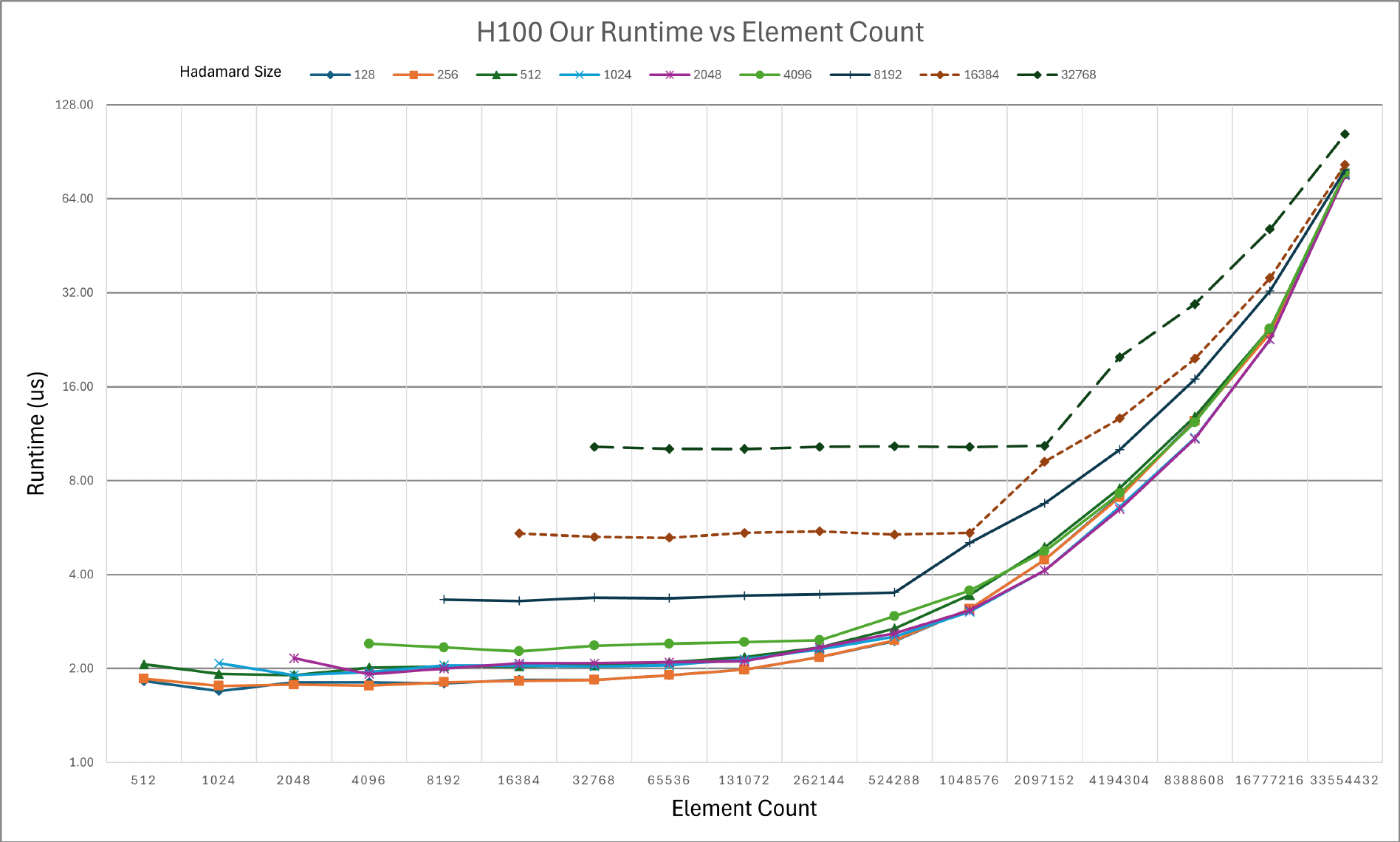}
    \includegraphics[width=\linewidth]{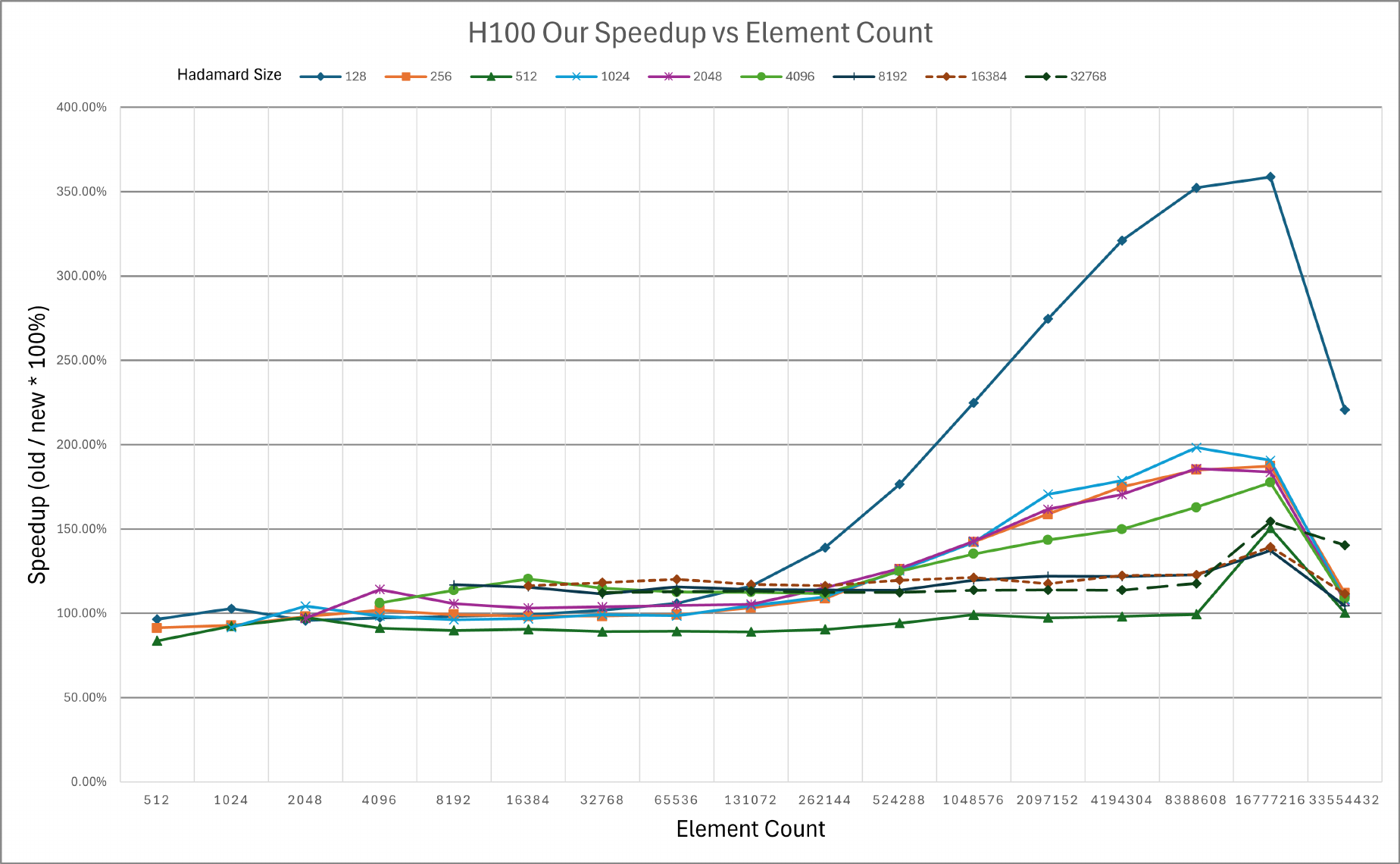}
    \caption{Graphs of runtime and speedup of HadaCore against the Dao AI Lab kernel, measured on an H100-PCIe. Each series represents a different rotation size, and the x-axis is the element count of the input tensor.}
\end{figure}

\subsection{Tables}

\begin{figure}[H]
    \centering
    \begin{subfigure}[b]{\linewidth}
        \includegraphics[width=\linewidth]{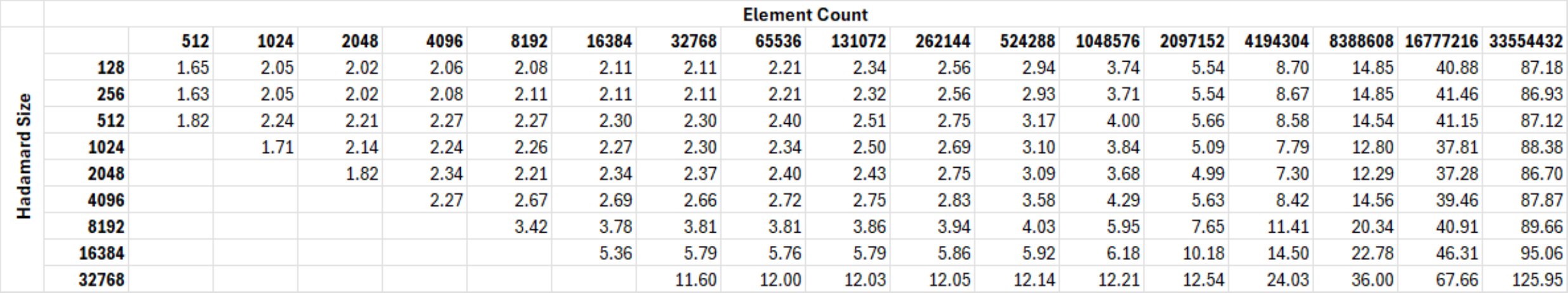}
        \caption{Runtime}
    \end{subfigure}
    \begin{subfigure}[b]{\linewidth}
        \includegraphics[width=\linewidth]{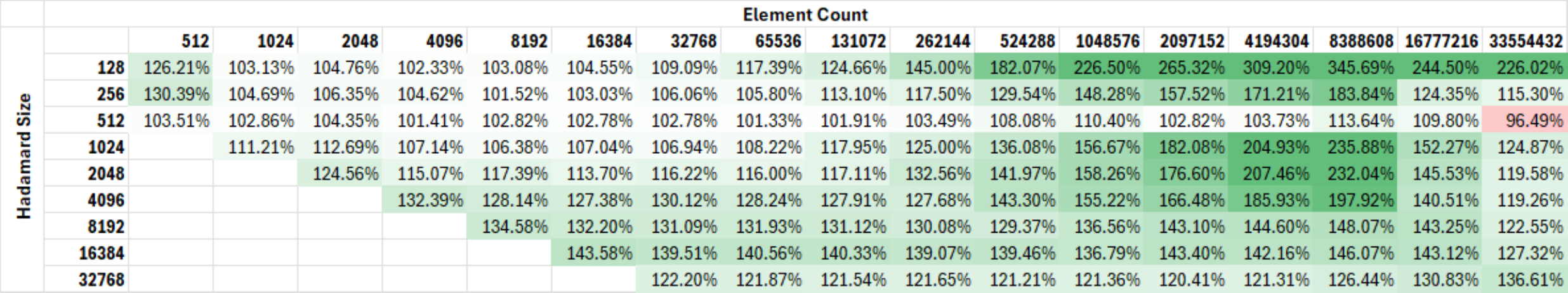}
        \caption{Speedup}
    \end{subfigure}
    \caption{Tables of runtime (in \textmu s) and speedup of HadaCore against the Dao AI Lab kernel, measured on an A100-PCIe.}
\end{figure}

\begin{figure}[H]
    \centering
    \begin{subfigure}[b]{\linewidth}
        \includegraphics[width=\linewidth]{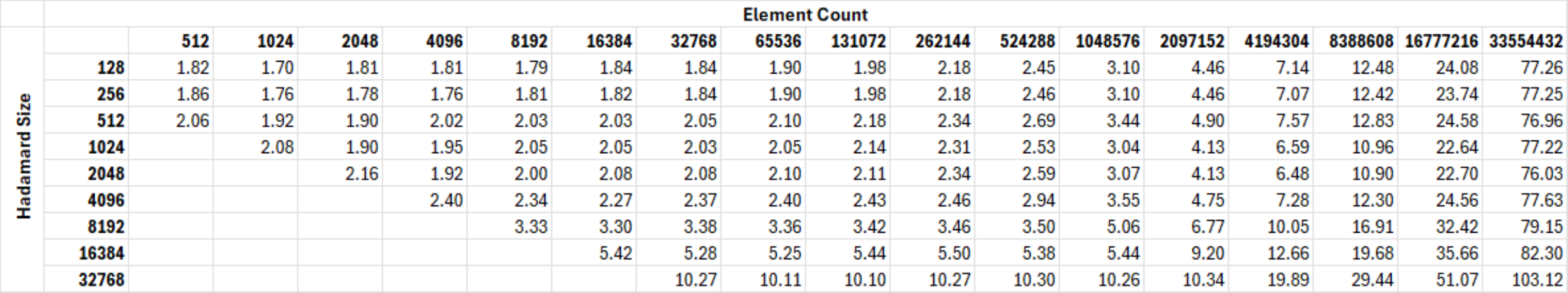}
        \caption{Runtime}
    \end{subfigure}
    \begin{subfigure}[b]{\linewidth}
        \includegraphics[width=\linewidth]{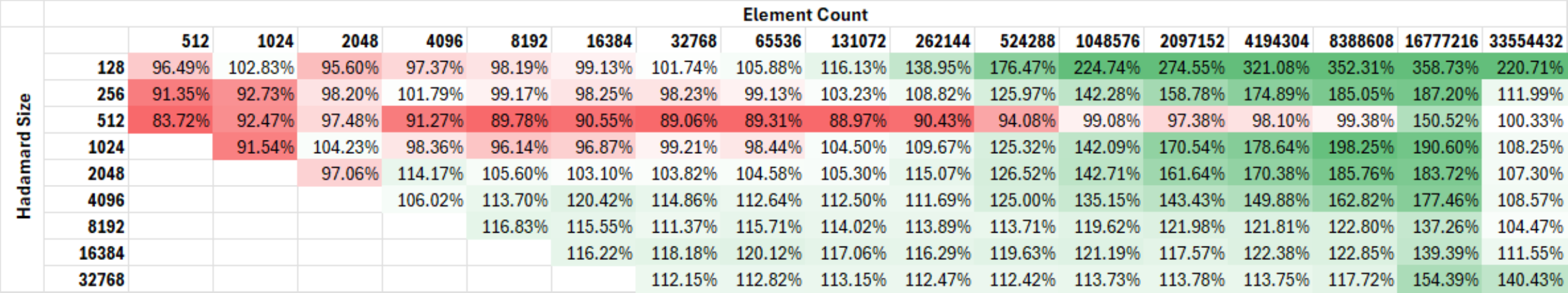}
        \caption{Speedup}
    \end{subfigure}
    \caption{Tables of runtime (in \textmu s) and speedup of HadaCore against the Dao AI Lab kernel, measured on an H100-PCIe.}
\end{figure}

\section{In-Place Rotation}
\label{appendix:inplace}

One simple optimization is modifying the input tensor in-place. If our tensor is 16M elements of FP16 (e.g. $4096\times 4096$), it will fit in \textasciitilde 32MB of cache. However, if we have a separate destination tensor, we need double that, \textasciitilde 64MB. The H100, A100, and L40S have 50MB, 40MB, and 48MB of L2 cache respectively, so for these enterprise GPUs, the source and destination tensors will evict each other’s lines from cache. While this analysis should theoretically only apply for sizes larger than half the L2 cache size but less than the full L2 cache size (like 32MB), in practice it might be different depending on the eviction policy and other memory usage.

This optimization can be applied to the \texttt{fast-hadamard-transform} library with a small change in \texttt{/csrc/fast\_hadamard\_transform.cpp}:

\begin{framed}
\begin{lstlisting}[language=c++]
- at::Tensor out = torch::empty_like(x);
+ at::Tensor out = x; // torch::empty_like(x);
\end{lstlisting}
\end{framed}
We verified that the result does not change empirically. The \texttt{fast-hadamard-transform} library now outperforms memcpy (which they used as a lower bound in their benchmarks) in some cases.

\begin{figure}[H]
    \centering
    \includegraphics[width=\linewidth]{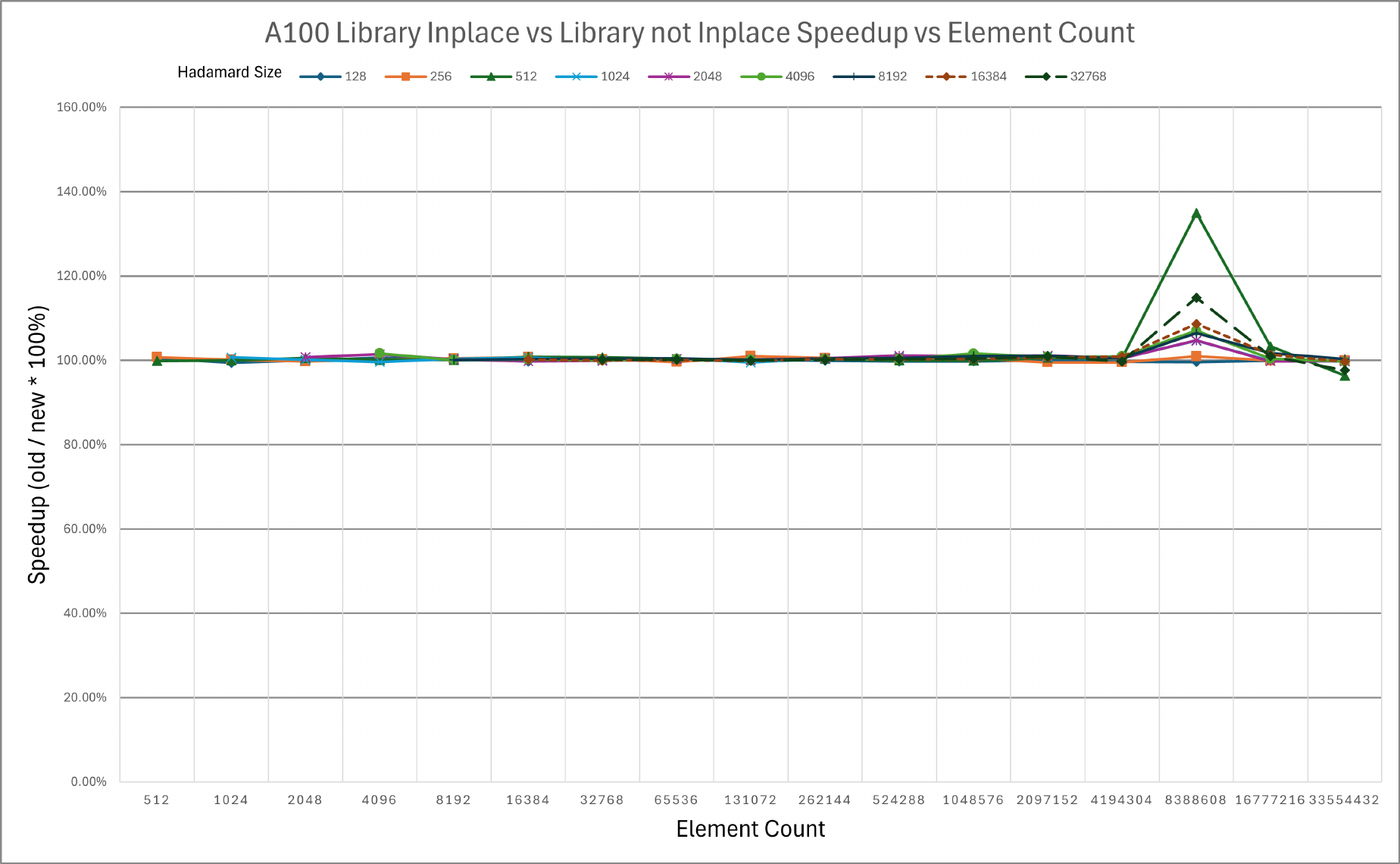}
    \includegraphics[width=\linewidth]{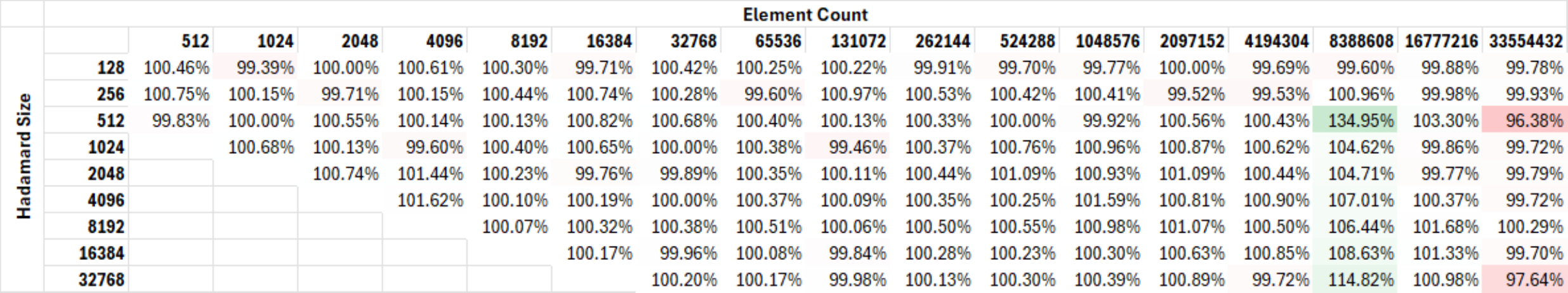}
    \caption{Graph and table of speedup when the Dao AI Lab \texttt{fast-hadamard-transform} is modified to perform the operation in-place, measured on an A100-PCie.}
\end{figure}

\raggedbottom

\begin{figure}[H]
    \centering
    \includegraphics[width=\linewidth]{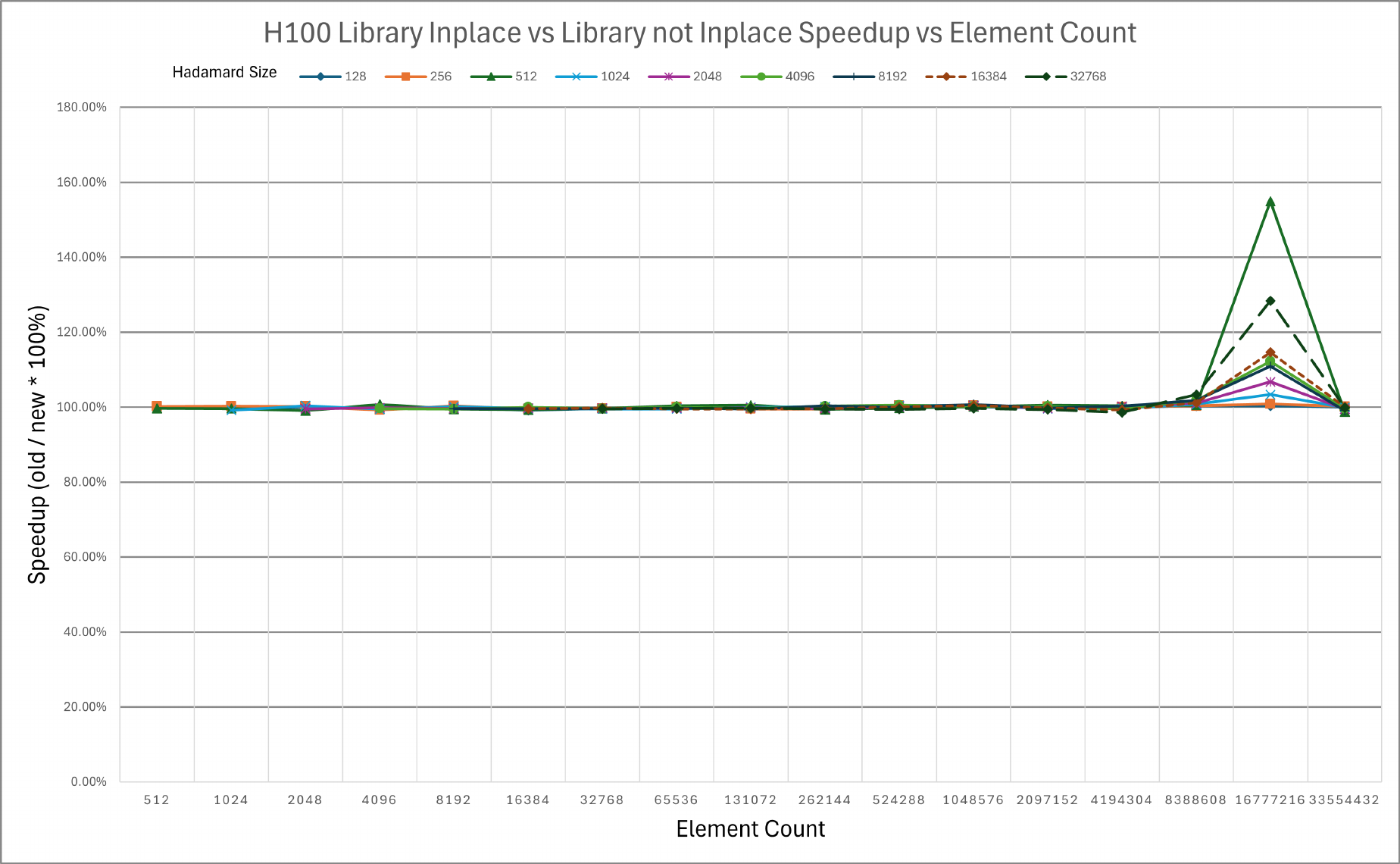}
    \includegraphics[width=\linewidth]{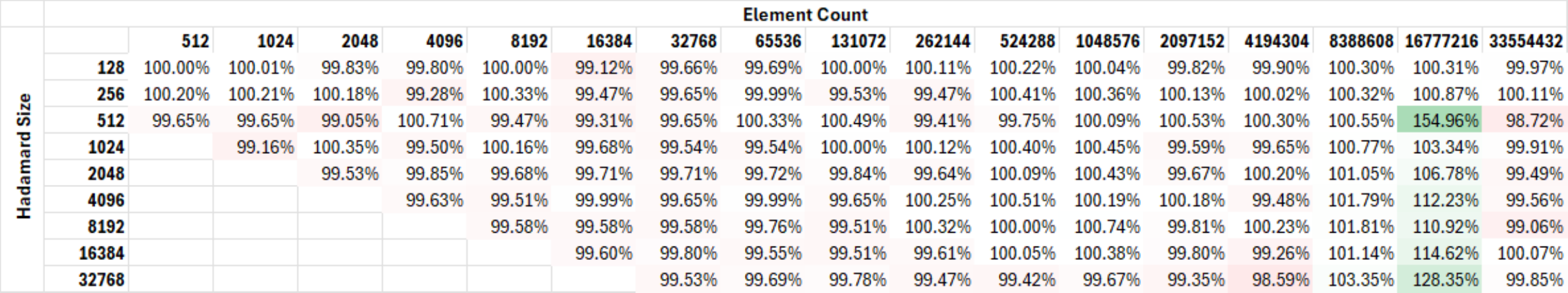}
    \caption{Graph and table of speedup when the Dao AI Lab \texttt{fast-hadamard-transform} is modified to perform the operation in-place, measured on an H100-PCie.}
\end{figure}

\section{BF16 Support}

In addition to FP16, we add BF16 support to our kernel. $16\times 16$ by $16\times 8$ \texttt{mma} is supported for BF16 on modern Nvidia GPUs, so we do not need to modify our strategy for performing the Hadamard transform with Tensor Cores to support BF16. The only difference is that BF16 Tensor Core instructions only allow accumulating the result in FP32, whereas FP16 Tensor Cores allow accumulating back into FP16 (which is what we use). To get around this, we use the FP32 accumulation but then immediately use native FP32 to BF16 type conversion instructions to get back a result in BF16. The added conversion instructions introduce an overhead that makes our BF16 kernel slightly slower than FP16, although we still demonstrate similar speedups as before when comparing our BF16 implementation to the Dao AI Lab's BF16 kernel. Our speedups are shown in the figures below for Nvidia A100 and H100 GPUs.

\begin{figure}[htbp]
    \centering
    \includegraphics[width=\linewidth]{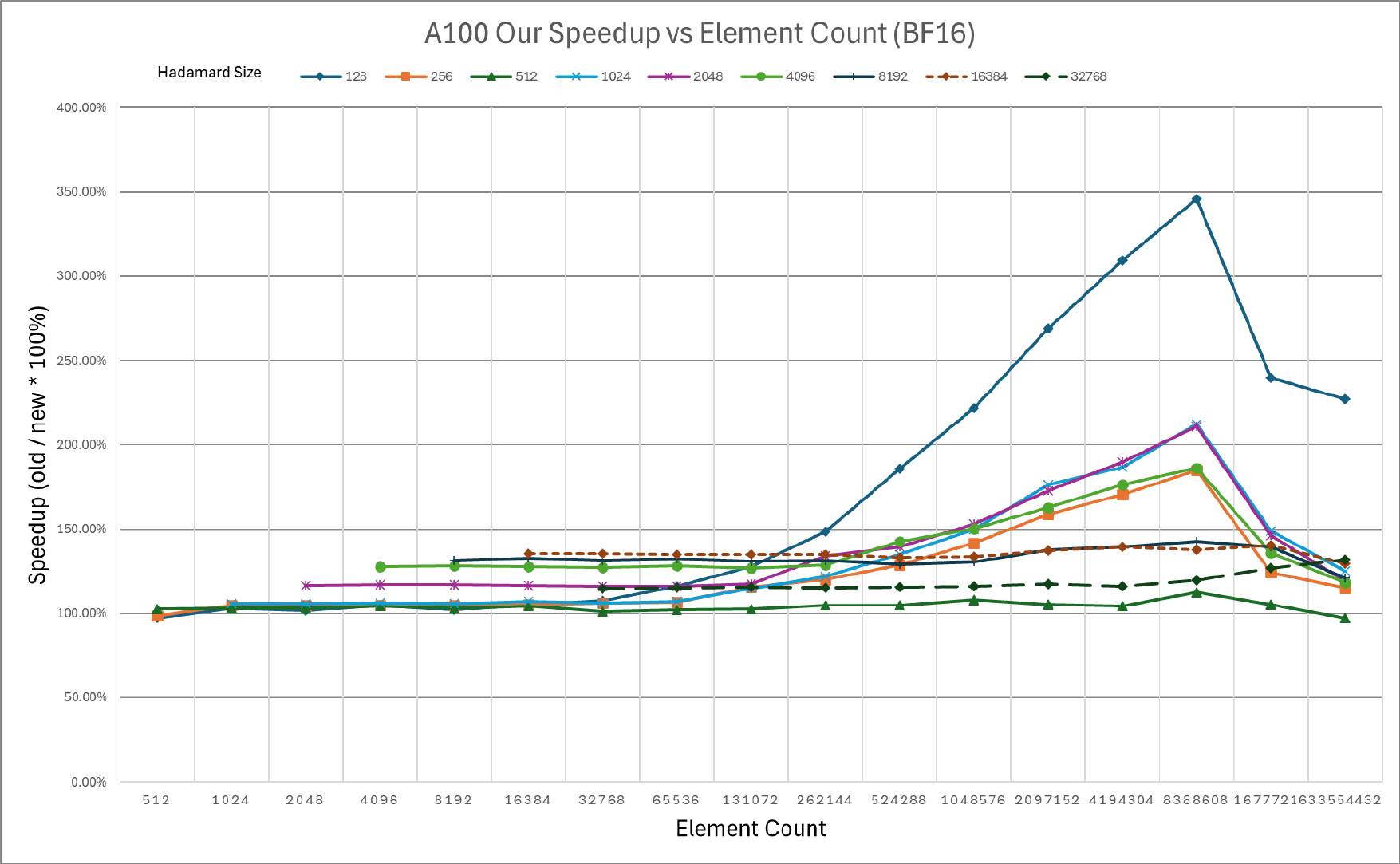}
    \includegraphics[width=\linewidth]{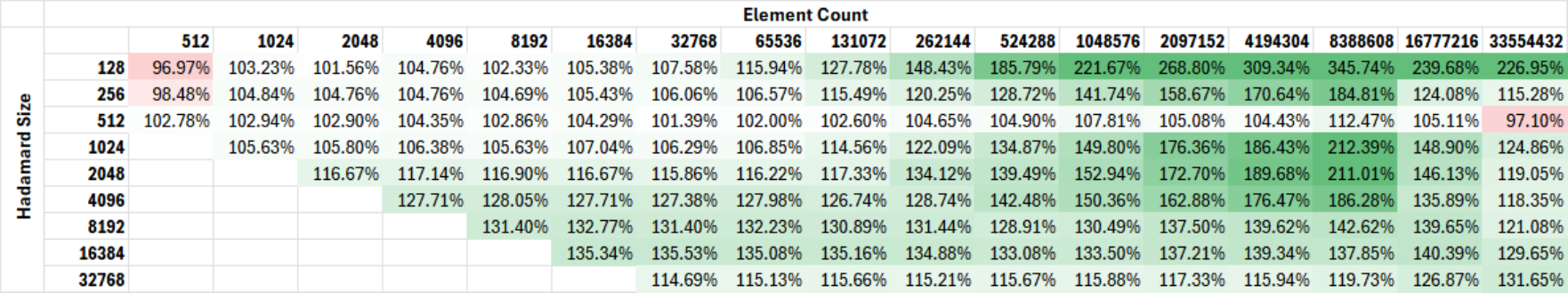}
    \caption{Graph and table of speedup of HadaCore against the Dao AI Lab kernel for BF16, measured on an A100-PCIe.}
\end{figure}
\begin{figure}[htbp]
    \centering
    \includegraphics[width=\linewidth]{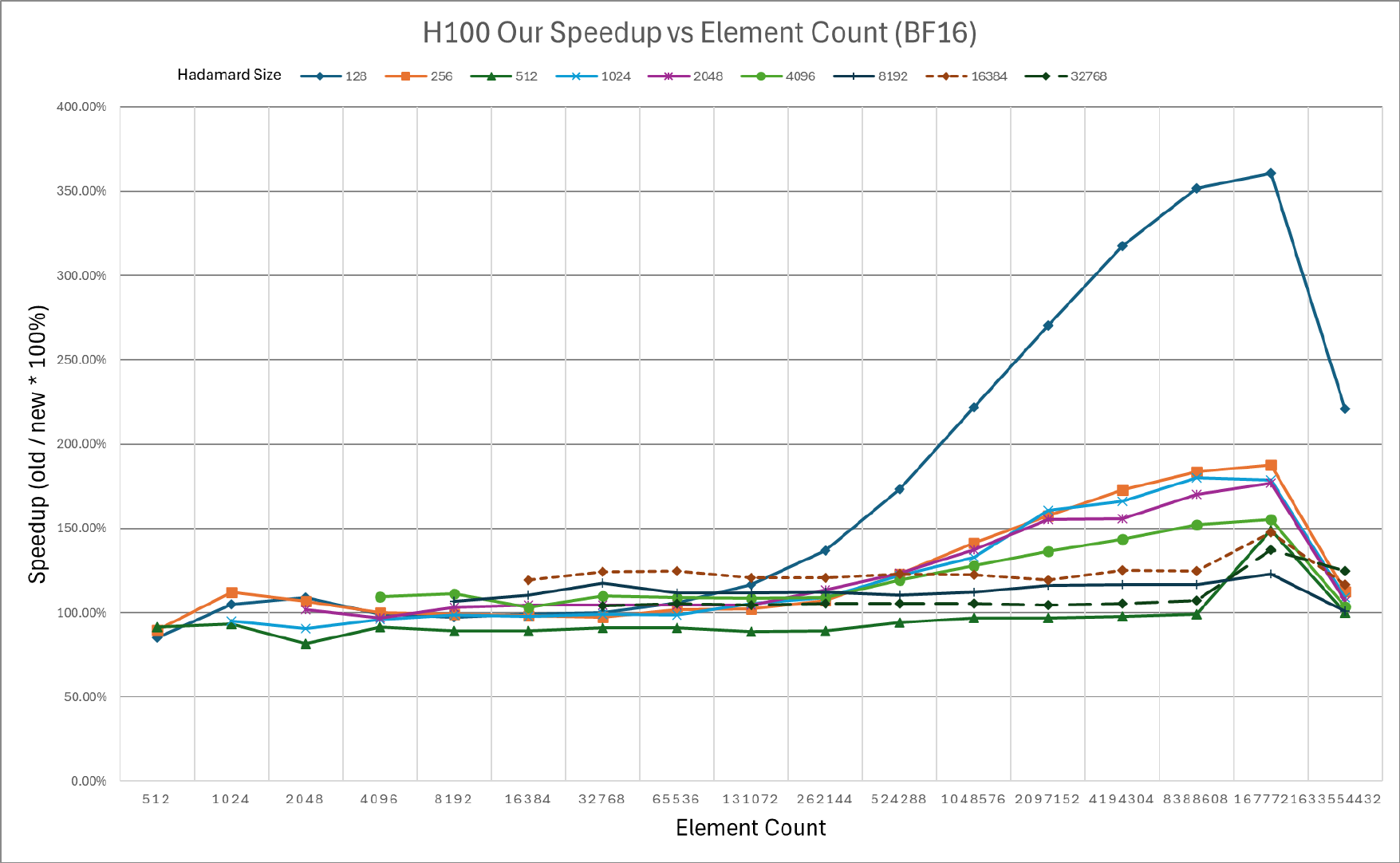}
    \includegraphics[width=\linewidth]{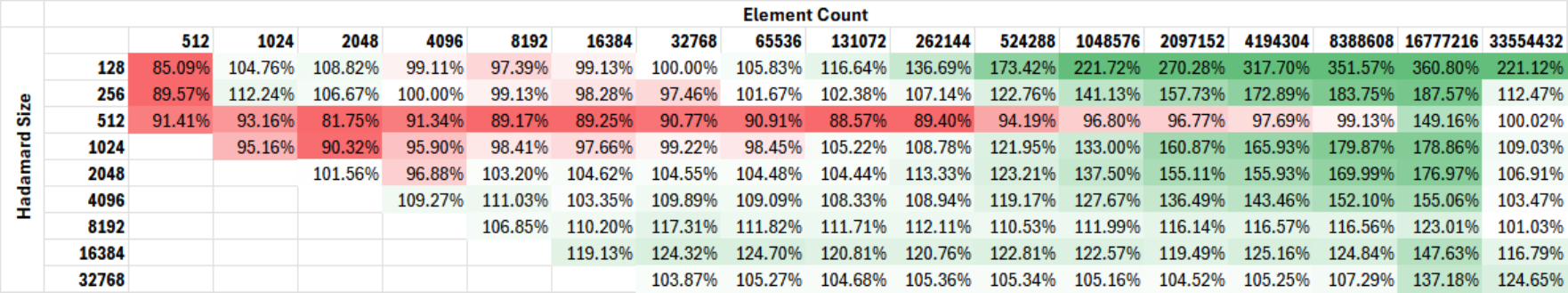}
    \caption{Graph and table of speedup of HadaCore against the Dao AI Lab kernel for BF16, measured on an H100-PCIe.}
\end{figure}

\end{document}

%% file: code/fwht.tex
\begin{framed}
\vspace{-2mm}
\begin{lstlisting}[language=Python]
def fwht(a) -> None:
    # In-place Fast Walsh-Hadamard Transform of array a.
    # Written by Wikipedia Contributors.
    h = 1
    while h < len(a):
        # perform FWHT
        for i in range(0, len(a), h * 2):
            for j in range(i, i + h):
                x = a[j]
                y = a[j + h]
                a[j] = x + y
                a[j + h] = x - y
        # normalize and increment
        a /= math.sqrt(2)
        h *= 2
\end{lstlisting}
\vspace{-2mm}
\end{framed}

%% file: references.bib
@misc{ashkboos2024quarotoutlierfree4bitinference,
      title={QuaRot: Outlier-Free 4-Bit Inference in Rotated LLMs}, 
      author={Saleh Ashkboos and Amirkeivan Mohtashami and Maximilian L. Croci and Bo Li and Pashmina Cameron and Martin Jaggi and Dan Alistarh and Torsten Hoefler and James Hensman},
      year={2024},
      eprint={2404.00456},
      archivePrefix={arXiv},
      primaryClass={cs.LG},
      url={https://arxiv.org/abs/2404.00456}, 
}

@misc{liu2024spinquantllmquantizationlearned,
      title={SpinQuant: LLM quantization with learned rotations}, 
      author={Zechun Liu and Changsheng Zhao and Igor Fedorov and Bilge Soran and Dhruv Choudhary and Raghuraman Krishnamoorthi and Vikas Chandra and Yuandong Tian and Tijmen Blankevoort},
      year={2024},
      eprint={2405.16406},
      archivePrefix={arXiv},
      primaryClass={cs.LG},
      url={https://arxiv.org/abs/2405.16406}, 
}

@misc{shah2024flashattention3fastaccurateattention,
      title={FlashAttention-3: Fast and Accurate Attention with Asynchrony and Low-precision}, 
      author={Jay Shah and Ganesh Bikshandi and Ying Zhang and Vijay Thakkar and Pradeep Ramani and Tri Dao},
      year={2024},
      eprint={2407.08608},
      archivePrefix={arXiv},
      primaryClass={cs.LG},
      url={https://arxiv.org/abs/2407.08608}, 
}

@ARTICLE{fastwalshhadamardtransform,
  author={Fino and Algazi},
  journal={IEEE Transactions on Computers}, 
  title={Unified Matrix Treatment of the Fast Walsh-Hadamard Transform}, 
  year={1976},
  volume={C-25},
  number={11},
  pages={1142-1146},
  doi={10.1109/TC.1976.1674569}}

@article{Sylvester1867LXTO,
  title={LX. Thoughts on inverse orthogonal matrices, simultaneous signsuccessions, and tessellated pavements in two or more colours, with applications to Newton's rule, ornamental tile-work, and the theory of numbers},
  author={James Sylvester},
  journal={Philosophical Magazine Series 1},
  year={1867},
  volume={34},
  pages={461-475},
  url={https://api.semanticscholar.org/CorpusID:118420043}
}

@misc{tridaofht,
  author = {Tri Dao},
  title = {Fast Hadamard Transform in CUDA, with a PyTorch interface},
  year = {2024},
  publisher = {GitHub},
  journal = {GitHub repository},
  howpublished = {\url{https://github.com/Dao-AILab/fast-hadamard-transform}},
  commit = {d1a56eee9d502e67faacf61b7b947180d66b32a0}
}

@misc{tensorcore,
  author = {Nvidia Corporation},
  title = {CUDA C++ Programming Guide},
  year = {2024},
  howpublished = {\url{https://docs.nvidia.com/cuda/cuda-c-programming-guide/index.html\#warp-matrix-functions}},
}

@misc{triton,
  author = {Philippe Tillet},
  title = {CUDA C++ Programming Guide},
  year = {2021},
  howpublished = {\url{https://openai.com/index/triton/}},
}

@misc{hendrycks2021measuringmassivemultitasklanguage,
      title={Measuring Massive Multitask Language Understanding}, 
      author={Dan Hendrycks and Collin Burns and Steven Basart and Andy Zou and Mantas Mazeika and Dawn Song and Jacob Steinhardt},
      year={2021},
      eprint={2009.03300},
      archivePrefix={arXiv},
      primaryClass={cs.CY},
      url={https://arxiv.org/abs/2009.03300}, 
}

@misc{tseng2024quipbetterllmquantization,
      title={QuIP\#: Even Better LLM Quantization with Hadamard Incoherence and Lattice Codebooks}, 
      author={Albert Tseng and Jerry Chee and Qingyao Sun and Volodymyr Kuleshov and Christopher De Sa},
      year={2024},
      eprint={2402.04396},
      archivePrefix={arXiv},
      primaryClass={cs.LG},
      url={https://arxiv.org/abs/2402.04396}, 
}
